\documentclass{aa}
\usepackage{txfonts}
\usepackage{natbib}

\usepackage{epsfig}
\usepackage{graphicx}

\usepackage{epstopdf}
\newcommand{\eqref}[1]{(\ref{#1})}

\begin{document}

\title{The effect of the environment on the P1/P2 period ratio for kink oscillations of coronal loops}
\author{B. Orza\inst{1} \and I. Ballai\inst{1}  \and R. Jain\inst{1} \and K. Murawski\inst{2}}
\institute{Solar Physics and Space Plasma Research Centre (SP$^2$RC), Department of
Applied Mathematics, The University of Shef{}field, Shef{}field,
UK, S3 7RH, email: {\tt\{b.orza;i.ballai;r.jain\}@sheffield.ac.uk}
\and Institute of Physics, Marie Curie-Sklodowska University, ul. Radziszewskiego 10, 20-031 Lublin, Poland, email: {\tt kmur@kft.umcs.lublin.pl}}
\date{Received dd mmm yyyy / Accepted dd mmm yyyy}

\authorrunning{Orza et al.}
\titlerunning{The effect of the environment on the $P1/P2$ period ratio}

\abstract{}
{The $P_1/P_2$ period ratio of transversal loop oscillations is currently used for the diagnostics of longitudinal structuring of coronal loops as its deviation from 2 is intrinsically connected to the density scale-height along coronal loops and/or the sub-resolution structure of the magnetic field. The same technique can be applied not only to coronal structures, but also to other oscillating magnetic structures.}
{The oscillations in magnetic structures are described by differential equations whose coefficients depend on the longitudinal structure of the plasma. Using a variational principle written for the transversal component of the velocity vector, developed earlier by McEwan et al. (2008), we investigate how the different temperature of the environment compared to the temperature of the magnetic structure will influence the $P_1/P_2$ ratio for typical coronal and prominence conditions. The possible changes are translated into quantities that are used in the process of remote plasma diagnostics in the solar atmosphere.}
{Using a straightforward, yet comprehensive, procedure we show that under coronal conditions the effect of the temperature difference between the plasma inside and outside the magnetic structure can change considerably the period ratio; in the case of coronal loops the change in the period ratio can reach even 40\%. We also show that once dispersive effects are taken into account, with oscillation periods shorter than the cut-off period (determined by the density and temperature difference) the domain where the model can be applied is reduced. In the case of prominences embedded in the hot corona, the effect of the environment is negligeable given the high density and temperature difference between the chromospheric prominences and corona. Using a numerical approach, we show that our analytical discussion produces a robust result. We also discuss what implications our model has on seismological (or diagnostics) techniques in the solar corona.  }
{Our analysis shows that the period ratio $P_1/P_2$ is sensitive to the temperature difference between the loop and its environment and this effect should always be taken into account when estimating the degree of density structuring with period ratio method of coronal loops.}


\maketitle

\section{Introduction}

Sudden energy releases in the solar atmosphere are known to generate
large scale global waves propagating over long distances (see, e.g. Moreton and Ramsey 1960, Uchida 1970, Thompson et al. 1999, Ballai et al. 2005). The energy
stored in these waves can be released by traditional dissipative
mechanisms, but it could also be transferred to
magnetic structures which may come in contact with global waves. This scenario is true not only for coronal structures, but applies to all magnetic entities in the solar atmosphere that can serve as waveguides (see, e.g. Wills-Davey and Thompson 1999, Patsourakos and Vourlidas 2009, Liu et al. 2010, etc).
In the corona, EIT waves generated by coronal mass ejections (CMEs)
and/or flares could interact with coronal loops, resulting in
the generation of kink modes, i.e. oscillations which exhibit
periodic movement about the loop's symmetry axis. Global waves can also interact with prominence fibrils as observed by, e.g. Ramsey and Smith (1966), and more recently by Eto et al. (2002), Jing et al. (2003), Okamoto et al. (2004), Isobe and Tripathi (2007), Pint\'er et al. (2008).
Oscillations of magnetic structures were (and are currently) used as a basic ingredient in one
of solar physics most dynamically expanding fields, namely {\it
coronal seismology}, where observations of wave characteristics (amplitude, wavelength, propagation speed, damping time/length) are corroborated with theoretical modelling (dispersion and evolutionary equations, as well as MHD models) in order to derive quantities that cannot be directly or indirectly measured (magnetic field magnitude and sub-resolution structuring, transport coefficients, heating functions, thermal state of the plasma, stratification parameters, etc.). Considerable advances have been achieved in diagnosing the state of the field and plasma (see, e.g.Roberts et al. 1984, Nakariakov et al. 1999, Nakariakov and Ofman 2001, Ofman and Thompson 2002, Ruderman and Roberts 2002, Andries et al. 2005, 2009, Ballai et al. 2005, 2011, Gruszecki et al. 2006, 2007, 2008, Banerjee et al. 2007, Ofman 2007, 2009, McLaughlin and Ofman 2008, Verth et al. 2007, Ballai 2007, Ruderman et al. 2008, Van Doorsselaere et al. 2008, Verth and Erd\'elyi 2008, Verth et al. 2008, Morton and Erd\'elyi 2009, Ruderman and Erd\'elyi 2009, Andries et al. 2009, Selwa and Ofman 2009, Selwa et al. 2010). It is highly likely that higher resolution observations made possible recently by space satellites such as STEREO, Hinode, SDO (and future missions) will further help understanding the complicated reality of the solar plasma environment. Indeed since their launch, data provided by these satellites are already shedding light on numerous aspects of coronal seismology, e.g. Verwichte et al. (2009) used STEREO data  to determine the three-dimensional geometry of the loop, SDO/AIA data was used by Aschwanden and Schrijver (2011) to prove the coupling of the kink mode and cross-sectional oscillations that could be explained as a consequence of the loop length variation in the vertical polarization mode. Finally, based on Hinode data, Ofman and Wang (2008) provided the first evidence for transverse waves in coronal multithreaded loops with cool plasma ejected from the chromosphere flowing along the threads. On the other hand the development of even the fundamental mode is not always guaranteed, as was shown observationally by, e.g. Aschnwanden et al. (2002) and later using MHD modelling by Selwa and Ofman (2010) and Selwa et al. (2011 a,b).

The dispersion
relations for many simple (and some quite complicated) plasma
waves under the assumptions of ideal magnetohydrodynamics (MHD)
are well known; they were derived long before accurate EUV
observations were available (see, e.g. Edwin and Roberts 1983, Roberts et al. 1984)
using simplified models within the framework of ideal and linear
MHD. Although the realistic interpretation of many observations is
made difficult especially by the poor spatial resolution of present
satellites not being quite sufficient, considerable amount of information
about the thermodynamical and dynamical state of the plasma, and the structure and magnitude of the
coronal magnetic field, can still be obtained.

The mathematical description of waves and oscillations in solar structures is, in general, given by equations whose coefficients vary in space and time. It has been recognised recently by, e.g.  Andries et al. (2005) that the longitudinal stratification (i.e. along the longitudinal symmetry axis of the tube that coincides with the direction of the magnetic field) modifies the periods of oscillations of coronal loops. Accordingly, in the case of kink waves, these authors showed that the ratio $P_1/P_2$ (where $P_1$ refers to the period of the fundamental transversal oscillation, while $P_2$ stands for the period of the first harmonic of the same oscillation) can differ - sometimes considerably - from the value of 2, that would be recovered if the loops were homogeneous. These authors also showed that the deviation of $P_1/P_2$ from 2 is proportional to the degree of stratification (see also, e.g. McEwan et al. 2006, Van Doorsselaere et al. 2007, Ballai et al. 2011). Later, studies by, e.g. Verth et al. (2007), showed that it is not only density stratification that is able to modify the  $P_1/P_2$ period ratio, but the variation of the loop's cross section area has also an effect on the period ratio. While the density stratification tends to decrease the period ratio, a modification of the cross section (i.e. when the magnetic field is expanding as we approach the apex) tends to increase the $P_1/P_2$ value. Observationally is much easier to detect the fundamental mode of kink oscillations (having the largest amplitude and the smallest damping rate), however high resolution observations made possible the evidence of even higher harmonics (see, e.g. De Moortel and Brady 2007, Van Doorsselaere et al. 2009).

In this study we will restrict our attention only to the effect of density stratification, the effect of the magnetic field structuring is left for a later analysis. The period ratio $P_1/P_2$ is connected to the density scale-height that quantifies the variation of density along the magnetic structure. All previous studies considered that the density stratification (indirectly the scale-height) is identical inside and outside the magnetic structure. However, the scale-height is directly linked to the temperature (via the sound speed) and an equal scale-height would mean an equal temperature, clearly not applicable for, e.g. coronal loops and/or prominence fibrils.

The aim of this paper is to investigate, using a simple mathematical approach, the effect of the environment on the period ratio $P_1/P_2$ and the consequences of the inclusion of a distinct environment  on estimations of the degree of density stratification. The paper is structured in the following way: in Section 2 we introduce the mathematical formalism and obtain analytical results for typical coronal and prominence conditions. Later, in Section 3 we use our findings to draw conclusions on the implications on coronal seismology and we present a method that could help diagnosing not only the degree of stratification in the loop, but also the temperature ratio of the plasma inside and outside the loop. Finally, our results are summarised in the last section.

\section{The mathematical formulation of the problem and analytical results}

EUV observations made by the recent high resolution space satellites (SOHO, TRACE, STEREO, Hinode, SDO) showed that, after all, coronal loops are enhancements of plasma (tracing the magnetic field in virtue of the frozen-in theorem) and the density of a typical loop can be as much as 10 times larger than the density of the environment. The heating of these coronal structures - according to the accepted theories (see, e.g. Klimchuk 2006, Erd\'elyi and Ballai 2007 and references therein)- occurs at the footpoints, while thermal conduction, flows, waves, instabilities and turbulences will help the heat to propagate along the full length of the loop. It is also obvious (as see in X-ray by, e.g. Hinode/XRT) that the temperature of the loop exceeds the temperature of the environment. A typical length of a coronal loop is 20-200 Mm, which means that the density inside the loop (seen in EUV) can vary by an order of magnitude, leading to the necessity of studying the effect of density stratification on the oscillations of coronal loops.

As pointed out for the first time by Dymova and Ruderman (2005, 2006), the propagation of kink waves in a straight tube in the thin tube approximation can be described by
\begin{equation}
\frac{\partial^2 v_r}{\partial t^2}-c_K^2(z)\frac{\partial^2 v_r}{\partial z^2}=0,
\label{eq:2.1}
\end{equation}
where $v_r$ denotes the radial (transversal) component of the velocity vector and the quantity $c_K$ is the propagation speed of kink waves (often called the density weighted Alfv\'en wave) defined as (see e.g. Edwin and Roberts 1983)
\begin{equation}
c_K=\left(\frac{\rho_iv_{Ai}^2+\rho_ev_{Ae}^2}{\rho_i+\rho_e}\right)^{1/2},
\label{eq:2.2}
\end{equation}
where $\rho_i$ and $\rho_e$ denote the densities inside and outside the coronal loop, and $v_{Ai}$ and $v_{Ae}$ represent the propagation speeds of the internal and external Alfv\'en waves, respectively and the dynamics is treated in the cold plasma approximation. For identical magnetic field inside and outside the tube, the fact that $\rho_i>\rho_e$ means that $v_{Ae}>v_{Ai}$. If we lift the thin flux tube restriction, then Eq. (\ref{eq:2.1}) must be complemented by terms that would describe dispersion. It is interesting to note that a similar equation was found recently by Murawski and Musielak (2010) describing Alfv\'en waves.
As we specified, our approach is using the cold plasma approximation in which the dynamics of kink oscillations in a coronal loop is described by Eq. (\ref{eq:2.1}).  For the sake of completeness we need to mention that the cold plasma approximations is not always true, especially in hot coronal loops observed by the SUMER instrument (e.g. Wang et al. 2003). It is very likely that the dynamics of kink oscillation will be described by a similar equations as Eq. (\ref{eq:2.1}) but with an extra term resulting from considering the effect of pressure perturbation. Furthermore the consideration of higher values of plasma-$\beta$ will affect the values of eigenfrequencies as was found by McLaughlin and Ofman (2004), and Ofman (2010, 2011a).

Assuming that all temporal changes occur with the same frequency, $\omega$, we can consider that the temporal dependence of variables (including $v_r$) has the form $\exp(i\omega t)$, which means that the PDE given by Eq. (\ref{eq:2.1}) transforms into
\begin{equation}
\frac{d^2v_r}{dz^2}+\frac{\omega^2}{c_K^2(z)}v_r=0.
\label{eq:2.3}
\end{equation}
 In reality the kink speed does not depend only on the longitudinal coordinate, $z$, but on all 3 coordinates. It is known that the dependence on the transversal coordinate, $r$, leads to the phenomena requiring short transversal length scales (resonant absorption, phase mixing, turbulence, wave leakage), used to explain the rapid damping of kink oscillations (see, e.g. Ofman and Aschwanden 2002, Ruderman and Roberts 2002, Ruderman 2008, etc). Equation (\ref{eq:2.3}) implies that the eigenfunctions, $v_r$, are driven by particular forms of $c_K(z)$, through the particular profile of the quantities that make up the kink speed (density, magnetic field). Inspired from the eigenvalue problem of Rayleigh-Ritz procedure, McEwan et al. (2008) used a variational principle that allows the calculation of eigenvalues, $\omega$, - a method that is employed by our analysis. Let us multiply the above equation by $v_r$ and integrate from the apex to the footpoint of the loop as
\begin{equation}
\int_0^Lv_r\frac{d^2v_r}{dz^2}dz+\omega^2\int_0^L\frac{v_r^2}{c_K^2(z)}dz=0.
\label{eq:2.4}
\end{equation}
Using integration by parts in the first integral (taking into account that for the fundamental mode $v_r(L)=dv_r(0)/dz=0$ and for the first harmonic $v_r(0)=v_r(L)=0$), the above equation simplifies to
\begin{equation}
\omega^2\int_0^L\frac{v_r^2}{c_K^2(z)}dz=\int_0^L\left(\frac{dv_r}{dz}\right)^2dz,
\label{eq:2.5}
\end{equation}
which results into the equation derived earlier by McEwan et al. (2008)
\begin{equation}
\omega^2=\frac{\Psi_1}{\Psi_2},
\label{eq:2.6}
\end{equation}
where
\[
\Psi_1=\int_0^L\left(\frac{dv_r}{dz}\right)^2dz, \quad \Psi_2=\int_0^L\frac{v_r^2}{c_K^2(z)}dz.
\]
In order to express the eigenvalue of such problem, we consider some trial functions for $v_r$ that satisfy the boundary conditions imposed at the footpoints and the apex of the loop. Since we are interested only in the characteristics of fundamental mode of kink oscillations and its first harmonic, we will assume that $v_r(z)$ will be proportional to $\cos(\pi z/2L)$ for the fundamental mode and $\sin(\pi z/L)$ for the first harmonic. It is obvious that these choices for eigenfunctions correspond to the homogeneous plasma, however - as we show in the Appendix - the corrections to the eigenfunction due to density stratification are rather small.

The problem of how the kink speed depends on the longitudinal coordinate, $z$, is a rather delicate problem and only simplified cases can be solved analytically. For simplicity, let us consider that the magnetic field inside and outside of the coronal loop are identical and homogeneous, while the density varies exponentially according to
\[
\rho_i(z)=\rho_i(0)\exp\left(\frac{2L}{\pi H_i}\sin\frac{\pi z}{2L}\right),
\]
\begin{equation}
 \rho_e(z)=\rho_e(0)\exp\left(\frac{2L}{\pi H_e}\sin\frac{\pi z}{2L}\right),
\label{eq:2.7}
\end{equation}
where $\rho_i(0)$ and $\rho_e(0)$ are the densities inside and outside the loop at $z=0$, i.e. at the the loop apex and $H_i$ and $H_e$ are the density scale-heights inside and outside the loop. Obviously the choice of density reflects a simplified description of the coronal loop model where plasma is isothermal and other further effects are neglected, however, this density profile allows us to obtain analytical results. A realistic description would require taking into account that the plasma is not isothermal (inside and outside the loop), the loop is curved and the density can depend on other coordinates, as well. This form of density dependence on the $z$ coordinate was earlier used by, e.g. Verth et al. 2007, McEwan et al. 2008, Morton and Erd\'elyi 2009, Morton and Ruderman 2011, Morton et al. 2011. With our chosen density profiles, the kink speed given by Eq. (\ref{eq:2.2}) becomes
\[
c_K^2(z)=\frac{2B_0^2}{\mu[\rho_i(z)+\rho_e(z)]}=
\]
\begin{equation}
=\frac{2v_{Ai}^2(0)}{\exp\left(\frac{2L}{\pi H_i}\sin\frac{\pi z}{2L}\right)+\xi^{-1}\exp\left(\frac{2L}{\pi H_i\chi}\sin\frac{\pi z}{2L}\right)},
\label{eq:2.8}
\end{equation}
where $B_0$ is the magnitude of the magnetic field, $v_{Ai}(0)$ is the Alfv\'en speed at the apex of the loop, and $\xi$ is the density ratio, i.e. $\rho_i(0)/\rho_e(0)$. Since the density outside the coronal loop is smaller than inside, we will consider that $\xi\geq 1$. The quantities $H_i$ and $H_e$ are the density scale-heights and they are proportional to the temperature of the plasma. Here we denoted $\chi=H_e/H_i$. Since the temperature of the loop is higher than its environment, we will take $\chi\leq 1$, so that the value $\chi=1$ corresponds to an identical density variation with height inside and outside the loop and identical temperatures, $\chi\rightarrow \infty$ resulting in a constant density in the environment of the loop, while the limit $\chi\rightarrow 0$ represents a case when the plasma inside the loop is homogeneous.

Using the particular form of $v_r$ for the fundamental mode and its first harmonic, we obtain that in the case of the fundamental mode
\[
\Psi_1^f=\frac{\pi^2}{8L},
\]
\[
\Psi_2^f=\frac{\pi}{8yv_{Ai}^2(0)}\times
\]
\begin{equation}
\times\left\{I_1(2y)+L_1(2y)+\xi^{-1}\chi\left[I_1\left(\frac{2y}{\chi}\right)+L_1\left(\frac{2y}{\chi}\right)\right]\right\},
\label{eq:2.10}
\end{equation}
where we introduced the dimensionless variable $y=L/\pi H_i$, $I_\nu(x)$ is the modified Bessel function of order $\nu$, $L_\nu(x)$ is the modified Struve function of order $\nu$, and the index $f$ stands for the fundamental mode. For the first harmonic we obtain that
\[
\Psi_1^1=\frac{\pi^2}{2L},
\]
\[
\Psi_2^1=\frac{\pi}{4y^2v_{Ai}^2(0)}\left\{\frac{3+2y^2}{y}\left[I_1(2y)+L_1(2y)\right]-3\left[I_0(2y)+\right]\right.
\]
\[
\left.+L_0(2y)+\frac{4y}{\pi}+\xi^{-1}\chi^2\left[\frac{\chi}{y}\left(3+\frac{2y^2}{\chi^2}\right)\left(I_1\left(\frac{2y}{\chi}\right)+L_1\left(\frac{2y}{\chi}\right)\right)\right.\right.
\]
\begin{equation}
\left.\left.-3\left(I_0\left(\frac{2y}{\chi}\right)+L_0\left(\frac{2y}{\chi}\right)\right)+\frac{4y}{\pi\chi}\right]  \right\},
\label{eq:2.11}
\end{equation}
where the superscript $1$ in the expressions of $\Psi_1^1$ and $\Psi_2^1$ stands for the first harmonic. Now using Eq. (\ref{eq:2.6}) for both modes we obtain that
\begin{equation}
\frac{P_1}{P2}=\sqrt{\frac{\Psi_2^f\Psi_1^1}{\Psi_1^f\Psi_2^1}}.
\label{eq:2.12}
\end{equation}
Inspecting the above relations we can see that the period ratio $P_1/P_2$ does not depend on Alfv\'en speed or loop length (they cancel out when calculating Eq. \ref{eq:2.12}). For coronal conditions we plot the period ratio given by Eq. (\ref{eq:2.12}) for $\xi=2$ with the variable $y$ varying between 0 and 10, although the larger values of $y$ are rather unrealistic since $y=10$ would correspond to a scale-height of 30 times shorter than the loop length (the scale-height corresponding to a typical temperature of 1 MK is 47 Mm).  Another variable in our problem is the ratio of scale-heights (i.e. temperatures), so $\chi$ will be varied in the interval 0 to 1.

The dependence of the $P_1/P_2$ period ratio on $\chi$ and the ratio $L/\pi H_i$ for coronal conditions ($\xi=2$) is shown in Fig. 1 with the case discussed earlier by, e.g. Andries et al. (2005) corresponding to the value $\chi=1$. In addition to the ratio $L/\pi H_i$ our model prescribes a possible diagnostic of the temperature difference between the loop and its environment.
\begin{figure}
\centering
\includegraphics[width=\columnwidth]{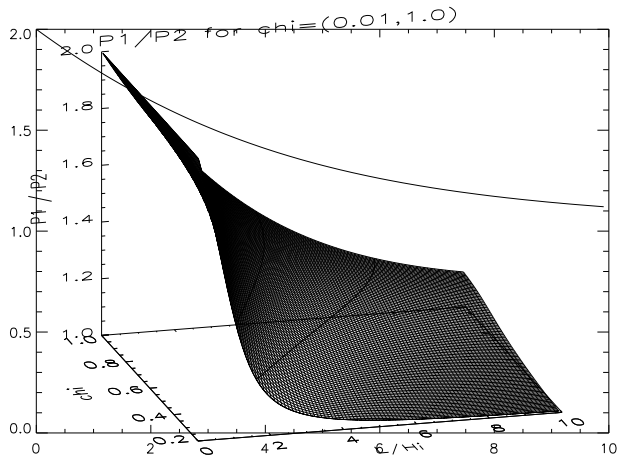}
\caption{The variation of the $P_1/P_2$ period ratio with the temperature parameter, $\chi$, and the ratio $L/\pi H_i$ for the case of a typical coronal loop (the density ratio, $\xi$, is 2). }
\label{fig1}
\end{figure}
The importance of changes when the different temperature of the environment is taken into account can be shown in a relative percentage plot shown in Figure 2.
\begin{figure}
\centering
\includegraphics[width=\columnwidth]{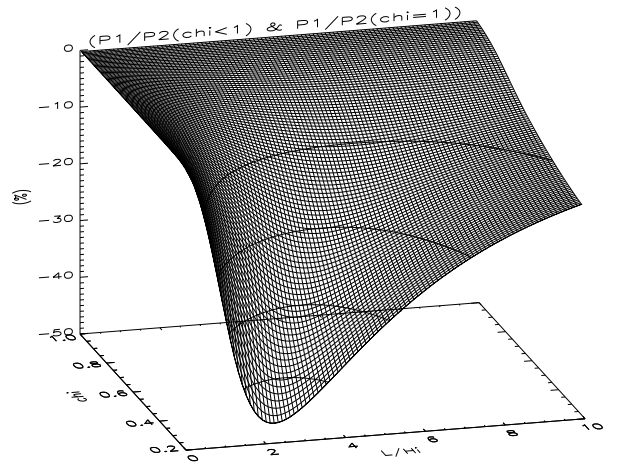}
\caption{The relative variation of the $P_1/P_2$ period ratio with the temperature parameter, $\chi$, and the ratio $L/H$ for the case of a typical coronal loop (the density ratio, $\xi$, is taken to be 2). }
\label{fig2}
\end{figure}
The relative change was calculated as the percentage change of the
results of our investigation compared to the case when $\chi=1$.
As we can see, the changes in the domain corresponding to values
of $\chi$ close to 1 are not significant. However, as the
temperature of the environment becomes lower than the temperature
inside the loop, this difference shows changes of the order of
10-20 \% for values of $\chi$ of up to 0.5, while for the cases
with $\chi$ near zero, the difference can be even 40 \% (for
$\chi=0.2$ and $L/\pi H_i=2$). Since the relative change is
negative, it means that for the same value of $P_1/P_2$ calculated
assuming the same temperature the ratio, $L/\pi H_i$ is
overestimated. A change of 25\% in the period ratio occurring at
approximative values of $L/\pi H_i=0.8$ and $\chi=0.65$ would mean
that for environment temperature that is 35 \% less than the loop
temperature, the scale-height is underestimated by about 25\%. It is important to note that the density ratio, $\xi$, does
have an important effect of the variation of period ratio. An
increase of $\xi$ to the value of 10 would result in relative
percentage change reduction and the maximum value of the change is
attaining its maximum value at 33 \% (for $\chi=0.2$ and $L/\pi
H_i=1$).

The same analysis was repeated for prominence structures. These structures are known to be of chromospheric origin and show rather long stability. Prominence fibrils are surrounded by much hotter and less denser corona. For these structure we suppose that the density of the prominence two orders of magnitude times higher, i.e. we take $\xi=100$. The typical temperature of prominences varies between $5\times 10^3$ and $10^4$ K, while the temperature of the surrounding corona can be even two orders of magnitude higher. That is why, the value of $\chi$ is chosen to change in the interval 50-150.
\begin{figure}
\centering
\includegraphics[width=\columnwidth]{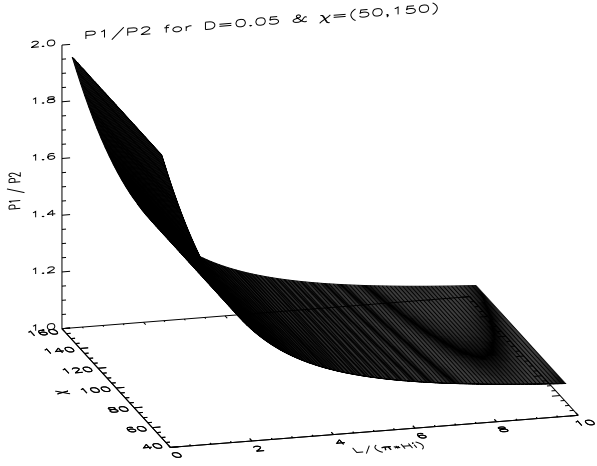}
\caption{The same as in Fig. 1 but we plot the variation of $P_1/P_2$ for prominences where $\xi=100$ and $\chi$ varies between 50 and 150. }
\label{fig3}
\end{figure}
As we can see in Fig. 3, the changes of the period ratio $P_1/P_2$ for prominences does not show large variation with $\chi$ and an analysis of the relative change (compared to the case corresponding to $\chi=50$) would reveal that these changes are of the order of 0.1 \%.

Strictly speaking the form of equilibrium densities given by Eq. (\ref{eq:2.7}) is obtained after imposing an equilibrium of forces along the vertical direction (considering that the loop is vertical) when the forces created by pressure gradients are balanced by gravitational forces. Moreover, the density scale-heights given before are connected to the gravitational acceleration. In an isothermal plasma the density scale-height is given as
\[
H=\frac{c_S^2}{\gamma g},
\]
where $c_S$ is the sound speed and $\gamma$ is the adiabatic index. In this case, the effect of the environment is described by a similar equation as given by Eq. (\ref{eq:2.1}), however, the governing equation is supplemented by an extra term that describes dispersive effects. Recently, Ballai et al. (2008) studied the nature of forced kink oscillations in a coronal loop and they obtained that the dynamics of the oscillations is given by an inhomogeneous equation of the form
\begin{equation}
\frac{\partial^2 v_r}{\partial t^2}-g\frac{\rho_i-\rho_e}{\rho_i+\rho_e}\frac{\partial v_r}{\partial z}-c_K^2(z)\frac{\partial^2 v_r}{\partial z^2}={\cal F},
\label{eq:2.14}
\end{equation}
 where ${\cal F}$ represents the external driver. This equation can be cast into a Klein-Gordon equation after introducing a new function, such that $v_r(z,t)=Q(z,t)e^{\lambda(z) z}$, where the value of $\lambda(z)$ is chosen in such a way that all first derivatives with respect to $z$ vanish. Introducing this ansatz into Eq. (\ref{eq:2.14}) we obtain
 \[
 \frac{\partial^2 Q}{\partial t^2}-\left[g\frac{\rho_i-\rho_e}{\rho_i+\rho_e}-2c_K^2(z)\Gamma(z)\right]\frac{\partial Q}{\partial z}+
 \]
 \[
+ \left[\frac{g(\rho_i-\rho_e)}{\rho_i+\rho_e}\Gamma(z)-c_K^2(z)\Gamma(z)^\prime-c_K^2(z)\Gamma(z)^2\right]Q-
\]
\begin{equation}
-c_K^2(z)\frac{\partial^2Q}{\partial z^2}={\cal F}e^{-\lambda z}.
 \label{eq:2.14.1}
 \end{equation}
 The condition that the coefficient of $\partial Q/\partial z$ is zero reduces to
 \begin{equation}
\Gamma(z)= \lambda^\prime z+\lambda=\frac{g(\rho_i-\rho_e)}{2c_K^2(z)(\rho_i+\rho_e)}.
 \label{eq:2.14.2}
 \end{equation}
With these restrictions, our governing equation transforms into
 \begin{equation}
 \frac{\partial^2 Q}{\partial t^2}-c_K^2\frac{\partial^2 Q}{\partial z^2}+\omega_C^2Q=e^{-\lambda z}{\cal F}.
 \label{eq:2.15}
 \end{equation}
 Here $\omega_C$ is the cut-off frequency of kink oscillations and is given by
 \[
 \omega_C^2=\left(\frac{2\pi}{P_C}\right)^2=
 \]
 \begin{equation}
 =\frac{g^2(\rho_i-\rho_e)^2}{4c_K^2(z)(\rho_i+\rho_e)^2}-\frac{gc_K^2(z)}{2}\frac{d}{dz}\left[\frac{\rho_i-\rho_e}{c_K^2(z)(\rho_i+\rho_e)}\right].
 \label{eq:2.16}
 \end{equation}
 It is obvious that the existence of this dispersive term is due to the different densities between the loop and its environment. The dispersion will affect the value of the $P_1/P_2$ period ratio. In the case of a homogeneous loop (understood in a local sense) where the propagation of kink oscillations is described by a Klein-Gordon equation, we can easily find that the period ratio is given by
 \begin{equation}
 \frac{P_1}{P_2}=2\left[1+\frac{Lg(1-\xi^{-1})}{\pi^2v_{Ai}^2}\right]^{-1/2},
 \label{eq:2.16.1}
 \end{equation}
 where the second term in the square bracket gives the deviation of $P_1/P_2$ from 2.

It is well known (see, e.g. Rae \& Roberts 1982, Ballai et al. 2006) that only those waves will be able to propagate in such magnetic structure whose frequencies are larger than the cut-off frequency given by Eq. (\ref{eq:2.16}). This condition will impose an upper boundary on the applicability region of the variables $L/\pi H_i$ and $\chi$.  The solution of Eq. (\ref{eq:2.15}) is a transversal kink oscillation propagating with speed $c_K$ followed by  a wake that is oscillating with the frequency $\omega_C$. Due to the height dependence of densities, the cut-off frequency will also depend on $z$. In terms of the variables used in our analysis, the cut-off frequency can be written as
 \[
 \omega_C=\frac{2\pi}{P_C}=\left\{\frac{g}{2(E_1+\xi^{-1}E_2)}\left[\frac{(gE_1-\xi^{-1}E_2)^2}{v_{A}^2(0)}-\right.\right.
 \]
 \begin{equation}
\left.\left.- \frac{\pi y}{L}\cos\frac{\pi z}{2L}(E_1-\frac{\xi^{-1}}{\chi}E_2)\right]\right\}^{1/2},
 \label{eq:2.16.2}
 \end{equation}
where we used the notations
\[
E_1=\exp\left(2y\sin\frac{\pi z}{2L}\right), \quad E_2=\exp\left(\frac{2y}{\chi}\sin\frac{\pi z}{2L}\right).
\]
In order to evaluate the effect of the cut-off on the period ratio, $P_1/P_2$, let us now return to Eq. (\ref{eq:2.14}) and repeat the same calculation as before. For simplicity we neglect the inhomogeneous part on the RHS of Eq. (\ref{eq:2.14}) and assume that perturbations oscillate with the the same real frequency, $\omega$. Using the same variational method, for the fundamental mode we obtain
 \begin{equation}
 \omega^2=\frac{\Psi_1^f+\Psi_1^{f\prime}}{\Psi_2^f}.
 \label{eq:2.17}
 \end{equation}
Here $\Psi_1^f$ and $\Psi_2^f$ are already defined by Eq. (\ref{eq:2.10}) and $\Psi_1^{f\prime}$ is simply given by
 \[
\Psi_1^{f\prime}=-\frac{g\pi^2}{8v_{Ai}^2(0)L}\left\{I_1(2y)+L_1(2y)-\frac{2(\xi-1)}{\pi\xi}-\right.
\]
\begin{equation}
\left.-\xi^{-1}\left[I_1\left(\frac{2y}{\chi}\right)+L_1\left(\frac{2y}{\chi}\right)\right]\right\}.
\label{eq:2.18}
\end{equation}
For the first harmonic we obtain that the frequency is given by
\begin{equation}
\omega^2=\frac{\Psi_1^1+\Psi_1^{1\prime}}{\Psi_2^1},
\label{eq:2.19}
\end{equation}
where $\Psi_1^1$ and $\Psi_2^1$ are specified by Eq. (\ref{eq:2.11}) and $\Psi_1^{1\prime}$ is defined as
\[
\Psi_1^{1\prime}=-\frac{\pi^2g}{4v_{Ai}^2(0)L}\left\{I_0(2y)+L_0(2y)-\frac{1}{y}(I_1(2y)+L_1(2y))\right.
\]
\begin{equation}
\left.+\frac{1}{\xi}\left[I_0\left(\frac{2y}{\chi}\right)+L_0\left(\frac{2y}{\chi}\right)-\frac{\chi}{y}\left(I_1\left(\frac{2y}{\chi}\right)+L_1\left(\frac{2y}{\chi}\right)\right)\right]\right\}.
\label{eq:2.20}
\end{equation}
The influence of the kink cut-off period on the period ratio $P_1/P_2$ is shown in Figs. 4 and 5 and its effect becomes obvious when Figs. 1 and 4 are compared.
\begin{figure}
\centering
\includegraphics[width=\columnwidth]{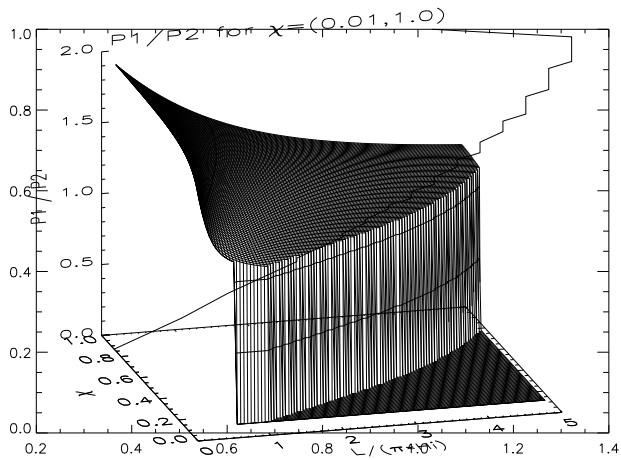}
\caption{The variation of $P_1/P_2$ period ratio for coronal loops when the effect of the cut-off period is taken into account .}
\label{fig5}
\end{figure}
First of all, we need to note that the effect of the dispersive term is more accentuated for increasing values of $L/\pi H_i$. Next, imposing the condition that the periods, we are investigating, are smaller than the cut-off period means that the domain of interest is restricted, as shown in Fig. 4 (in fact, only $P_1$ is required to be smaller than the cut-off period, as $P_2$ is always smaller than $P_1$). The region where the above condition is not satisfied was flagged by zero and the drop in the $P_1/P_2$ to zero represents the boundary of the region where this imposed condition is satisfied. Looking from above, the domain of applicability is shown in Fig. 5, with the domain labelled by index "I" indicating the region where the periods are smaller than the cut-off period, while the region "II" corresponds to the set of ($L/\pi H_i$, $\chi$) values for which no physical solution is found.
\begin{figure}
\centering
\includegraphics[width=\columnwidth]{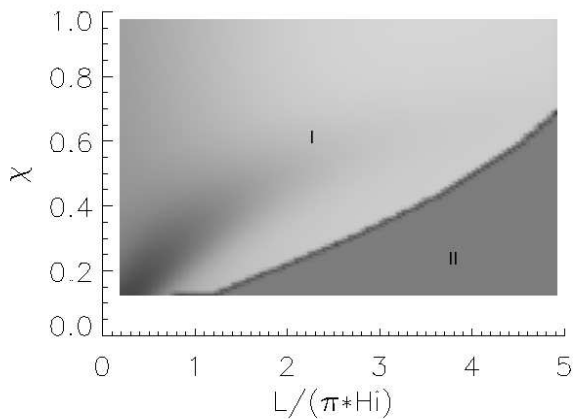}
\caption{The domain where the periods of fundamental mode and its first harmonic are smaller than the kink cut-off period, given by Eq. (\ref{eq:2.16}) for coronal conditions. The domain of permitted values is shown by label "I", while region "II" corresponds to the unphysical results. }
\label{fig6}
\end{figure}
When the dispersive term is not taken into account, the $P_1/P_2$ period ratio is independent on the length of the loop and the Alfv\'en speed (basically, the independent parameter is the density ratio, $\xi$). However, once the dispersive term is considered in Eq. (\ref{eq:2.14}), the $P_1/P_2$ ratio will depend on the length and the Alfv\'en speed measured at the apex of the loop. For the numerical example shown here, we have chosen a length of 150 Mm and an Alfv\'en speed at the apex of the loop ($z=0$) of 1000 km s$^{-1}$.  Since the length of the loop is given, varying $L/\pi H_i$ would mean a change in $H_i$. In our numerical analysis we stopped at $L/\pi H_i=5$ that corresponds to a scale-height of $9.5$ Mm. Assuming a loop in hydrostatic equilibrium, the value of $H=9.5$ Mm leads to a temperature of 0.2 MK. As the value of $L/\pi H_i$ decreases, the scale-height increases. It is easy to verify that for a fixed value of $L$, the $P_1/P_2$ period ratio is proportional to $v_{Ai}(0)$, and an increase/decrease of 200 km s$^{-1}$ would result in a change of only 4 \% towards the large $L/\pi H_i$ part of our investigated domain. If we fix the value of the Alfv\'en speed at the loop apex, the variation of $P_1/P_2$ is inversely proportional to $L$, but again, for a change of 50 Mm in $L$ drives changes of the order of 2\% for large values of $L/\pi H_i$.

For solar prominences we repeated the calculations but now assuming that $\xi=100$, $L=1$ Mm and $v_{Ai}(0)=120$ km s$^{-1}$ (see Fig. 6). Under these conditions it is obvious that possible solutions are found for $L/\pi H_i<2.4$.  The condition that a hydrostatic equilibrium is reached inside the prominence means that the smallest scale-height we use is $1.32\times 10^5$ m, which corresponds to a minimum temperature of 2800 K (nearly a quarter of the typical prominence temperature). As we approach smaller values for $L/\pi H_i$, the temperatures increase, so that when $L/\pi H_i=0.7$, the temperature is approximately $10^4$ K, a typical prominence temperature.
\begin{figure}
\centering
\includegraphics[width=\columnwidth]{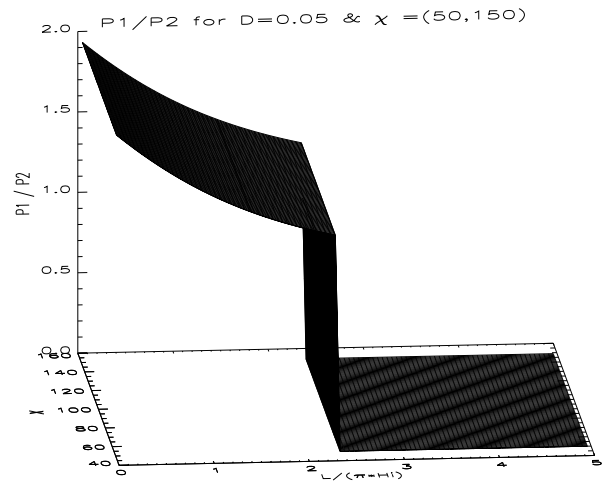}
\caption{The same as Figure 4, but now the period ratio is plotted for prominence conditions. }
\label{fig7}
\end{figure}

 \section{Implications for magneto-seismology}

The immediate implication of our calculations is that the $P_1/P_2$ period ratio has no one-to-one correspondence with the internal stratification of the magnetic structure, but depends also on the temperature ratio between the interior and exterior of the magnetic structure, i.e. an observed period ratio allows the diagnostics of the temperature ratio, too. Our analysis shows that the effect of temperature difference is more pronounced for those cases where the temperature inside the waveguide is larger than outside (e.g. coronal loops) and, in general, negligible in prominence cases. For coronal loops it is also evident that noticeable effects of the temperature difference on the $P_1/P_2$ ratio are encountered for relative small values of $L/\pi H_i$ (say, below 5) and temperature ratio that are smaller than 70\%.

Our analysis also opens a new way of diagnosing the multi-temperature loops and their environment. The relations derived in the present study show that the physical parameters entering the problem are the density ratio, temperature ratio and the ratio of loop length and scale-height. Out of these quantities, the density is the parameter that can be determined (although with errors) from emission, so we will suppose that the value of $\xi$ is known. The diagnostics of the loop in the light of the new introduced parameter becomes possible once we specify an additional relation connecting the temperature ratio and the density scale-height measured against the length of the loop. As a possibility we investigate the case when for the same loop we can determine now only the period of the fundamental mode ($P_1$) and its first harmonic ($P_2$) but also the period of the second harmonic, here denoted by $P_3$. Now we can form a new ratio, $P_1/P_3$, which can be determined in a similar way as above. Since the measurement of the three periods refer to the same loop we can estimate the value of $\chi$ and $L/\pi H_i$ in a very easy way.
\begin{figure}
\centering
\includegraphics[width=\columnwidth]{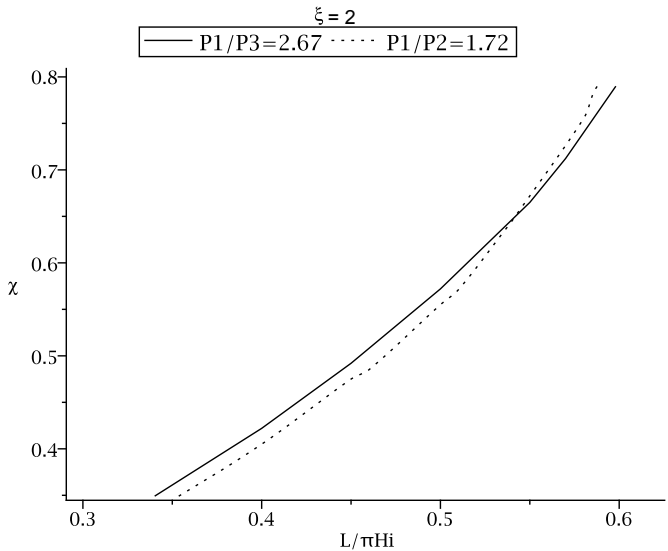}
\caption{An example on how the period ratio of the first three harmonics of a coronal loop kink oscillations can be used to diagnose the density scale-height of the loop and the temperature difference between the coronal loop and its environment. Here the $P_1/P_2$ dependence is shown by the dotted line while the solid line stands for the value of $P_1/P_3$.  }
\label{fig8}
\end{figure}
In Fig. 7 we illustrate such a case. We suppose that for a loop of half-length of 150 Mm with an Alfv\'en speed at the apex of the loop of 1000 km s$^{-1}$ we measure the period ratio of 1.72 for $P_1/P_2$ and 2.67 for $P_1/P_3$. By specifying the value of the period measurement means that in a dependence similar to the one shown in Fig. 1 we obtain an arc which in a ($L/\pi H_i, \chi$) coordinate system will look like the dotted line curve in Fig. 8. In a similar way, for the value of $P_1/P_3=2.67$ we would obtain another curve, here shown by the solid line (here the curves are the projections of the intersection of similar surfaces as in Fig 1 with the horizontal surface corresponding to the specified period ratio). Since the two measurements correspond to the same loop, their intersection point will give us the exact value of $L/\pi H_i$ and $\chi$. For the particular example used here we obtain $L/\pi H_i=0.57$ and $\chi=0.67$. Our results show very little sensitivity with the density ratio, for example, if the ratio would be 10 then the intersection point would change to $L/\pi H_i=0.54$ and $\chi=0.66$.

\section{Conclusions}

In order to carry out coronal seismology it is imperative to know
the relationship between the composition of a plasma structure and
the oscillations supported by the coronal loop. High resolution
observations make possible an accurate diagnostic of not only the
magnetic field strength, but also the thermodynamical state of the
plasma.

The period ratio $P_1/P_2$ (and its deviation for the canonical value of 2) belonging to the period of transversal fundamental kink mode and its first harmonic is a perfect tool for diagnosing the longitudinal structure of magnetic structures. In our study we investigated the effect of the environment on the period ratio assuming that the density scale-heights (implicitly the temperature) inside and outside magnetic structures are different.

Using a simple variational method first applied in this context by McEwan et al. (2008), we derived for the first time an analytical expression that connects the value of the period of kink oscillations and parameters of the loop. We showed that in the case of coronal loops the effect of temperature difference between the loop interior and exterior can lead to changes of the order of 30-40\% that could have significant implications on the diagnosis of longitudinal density structuring of the coronal loop. In the case of prominences, due to the very large density and temperature difference between the prominence and coronal plasma, the changes in $P_1/P_2$ due to the different temperature are very small. Once dispersive effects are taken into account (through a Klein-Gordon equation) the domain of applicability of $P_1/P_2$ seismology in the case of coronal loops becomes restricted and physically accepted solutions can not be found for any temperature ratio (here denoted by the parameter $\chi$).

Since our model introduces a new variable in the process of plasma diagnostic a new relation is needed that connects the parameters of interest. To illustrate the possibilities hidden in our analysis we have chosen the case when the same loop shows the presence of the additional second harmonic. Superimposing the dependences of the $P_1/P_2$ and $P_1/P_3$ ratios with respect to the temperature ratio factor, $\chi$ and $L/\pi H_i$ we could find the set of the values that satisfies a hypothetical measurement.

Finally we need to emphasise that our approach supposes a certain degree of simplification, therefore our results do not provide an absolute qualitative and quantitative conclusion. First we supposed that the loop is thin, and Eq. (\ref{eq:2.1}) can be applied to describe the dynamics of kink oscillations in coronal loops. It is obvious that this statement is not true for very short loops (the ratio of the loop ratio and its length is not very small) in which case, the governing equation has to be supplemented by an extra term. Secondly, our isothermal supposition of the loop and its environment is also that needs refinement as observations (see, e.g. Winebarger et al. 2003, Warren et al. 2008, Berger et al. 2011, Mulu-Moore 2011) show that the loops are not always in hydrostatic equilibrium nor isothermal. Here we supposed the idealistic situation of a static background, however recent analysis by Ruderman (2011) showed that the the temporal dependence of density through flow and cooling can also influence the ratio of the two periods.

\begin{acknowledgements}
IB acknowledges the financial support by NFS Hungary (OTKA, K83133). We are grateful to the anonymous Referee for his/her suggestions that helped to improve the quality of the paper.
\end{acknowledgements}

\appendix
\section{Corrections to the eigenfunctions due to the density stratification}
\begin{figure}
\includegraphics[width=\columnwidth]{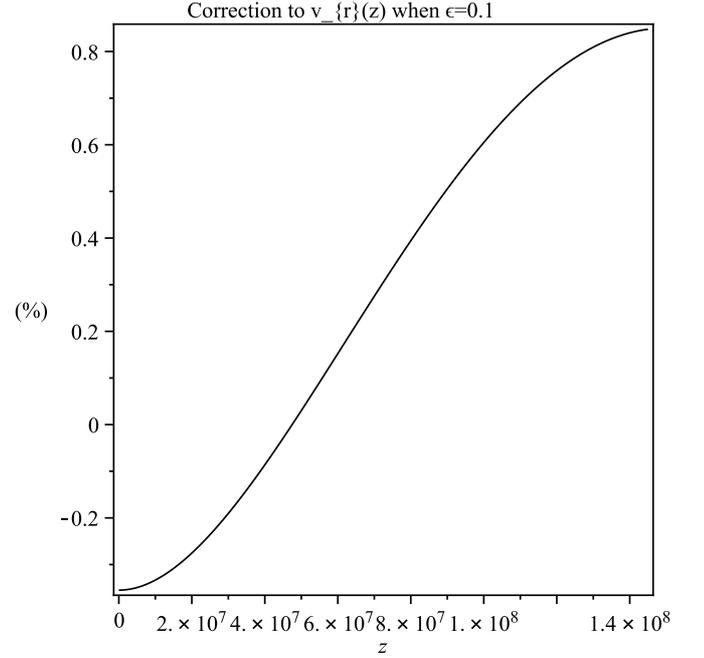}
\caption{ Correction to the eigenfunction for the fundamental mode kink oscillation when $\epsilon=0.1$. Here $L=1.5\times 10^8$ m represents the loop length}
\label{fig9}
\end{figure}

In the Appendix we estimate the corrections to the chosen eigenfunctions due to the density stratification. Analytical progress can be made in the small $y/\chi$ limit. Since $\chi$ is a value smaller than one, this condition would automatically mean that we work in the small $y$ limit, provided $\chi$ is not becoming too small.
We are interested only in the characteristics of fundamental mode of kink oscillations and its first harmonic. Following equation (\ref{eq:2.3}) with the boundary conditions $v_r(L)=dv_r(0)/dz=0$ and $v_r(0)=v_r(L)=0$ for the fundamental mode and first harmonic, we introduce a new variable so that
\[
 \zeta=\frac{z}{L}, \;\;\;\mbox{with}\;\;\; \frac{\omega L}{\sqrt{2}v_{Ai}}=\Omega.
\]
In the new notations the density inside the loop can be written as
\begin{equation}
\rho_i(\zeta)=\rho_i(0)\exp\left[\frac{h}{H_i}\sin\frac{\pi\zeta}{2}\right],
\label{eq:AA.1.1}
\end{equation}
where $h$ is the loop height above the solar atmosphere (a similar equation can be written for the external density). Next, we are working in the approximation $h/H_i=\epsilon\ll 1$, so equation (\ref{eq:2.3}) becomes
\begin{equation}\label{AA.1}
\frac{d^{2}v_{r}(\zeta)}{d\zeta^{2}}+\Omega^{2}\left[1+\epsilon sin\left(\frac{\pi \zeta}{2}\right)+\frac{1}{\xi}\left(1+\frac{\epsilon}{\chi}sin\left(\frac{\pi \zeta}{2}\right)\right)\right]v_{r}(\zeta)=0,
\end{equation}
where $\xi=\rho_{i}(0)/\rho_{e}(0)>1$ is the density ratio and $\chi=H_{e}/H_{i}<1$, with $H_{e}$ and $H_{i}$ being the density scale heights inside and outside the loop.
Let us we write $v_{r}$ and $\Omega$ as
\begin{equation}\label{AA.2}
v_{r}(\zeta)=v_{r}^{(0)}(\zeta)+\epsilon v_{r}^{(1)}(\zeta)+{\cal O}(\epsilon^2), \quad \Omega=\Omega_{0}+\epsilon \Omega_{1}+{\cal O}(\epsilon^2).
\end{equation}
Substituting these expansions into Eq. (\ref{AA.1}) and collecting terms proportional to subsequent powers of $\epsilon$ we obtain
\begin{equation}\label{eq:AA.4}
\frac{d^2v_{r}^{(0)}}{d\zeta^2}+(1+\xi^{-1})\Omega_{0}^2v_{r}^{(0)}=0,
\end{equation}
\[
\frac{d^2v_{r}^{(1)}}{d\zeta^2}+(1+\xi^{-1})\Omega_{0}^2v_{r}^{(1)}+2(1+\xi^{-1})\Omega_{0}\Omega_{1}v_{r}^{(0)}+
\]
\begin{equation}
+\Omega_{0}^2v_{r}^{(0)}\left(1+\frac{1}{\xi\chi}\right)sin\left(\frac{\pi \zeta}{2}\right)=0.
\label{AA.5}
\end{equation}
These equations must be solved separately for the fundamental mode and its first harmonic taking into account the boundary conditions
\begin{itemize}
\item{For the fundamental mode}
\[
v_{r}(\zeta=1)=0, \quad \frac{dv_{r}}{d\zeta}(\zeta=0)=0
\]
\item{For the first harmonic}
\[
v_{r}(\zeta=1)=v_r(\zeta=0)=0
\]
\end{itemize}
Let us first calculate the correction to the fundamental mode. It is easy to show that the solution of equation (\ref{eq:AA.4}) taking into account the above boundary condition becomes
\begin{equation}\label{AA.6}
v_{r}^{(0)}(\zeta)=\cos\left(\frac{\pi \zeta}{2}\right),
\end{equation}
and
\begin{equation}\label{AA.7}
\Omega_{0}=\frac{\pi}{2\sqrt{1+\xi^{-1}}}.
\end{equation}
It is important to note that the form of the above solution is exactly the same as the solution we employed for the eigenfunction, $v_r$. In the next order of approximation we obtain equation (\ref{AA.5}) which can be written as
\[
\frac{d^2v_{r}^{(1)}}{d\zeta^2}+(1+\xi^{-1})\Omega_{0}^2v_{r}^{(1)}=
\]
\begin{equation}
-2(1+\xi^{-1})\Omega_{0}\Omega_{1}v_{r}^{(0)}-\Omega_{0}^2v_{r}^{(0)}\left(1+\frac{1}{\xi\chi}\right)sin\left(\frac{\pi \zeta}{2}\right).
\label{eq:AA.7.1}
\end{equation}
This boundary value problem will permit solutions only if the right-hand side satisfies the compatibility condition that can be obtained after multiplying the left-hand side by the expression of $v_r^{(0)}$ and integrating with respect to the variable $\zeta$ between 0 and 1, or
\[
\int_{0}^{1}v_{r}^{(0)}\left[2(1+\xi^{-1})\Omega_{0}\Omega_{1}v_{r}^{(0)}+\Omega_{0}^2v_{r}^{(0)}\left(1+\frac{\xi^{-1}}{\chi}\right)sin\left(\frac{\pi \zeta}{2}\right)\right]d\zeta=0.
\]
After some straightforward calculus we can find that
\begin{equation}\label{AA.8}
\Omega_{1}=-\frac{1}{3(1+\xi^{-1})^{3/2}}\left(1+\frac{1}{\xi\chi}\right).
\end{equation}
As a result, equation (\ref{AA.5}) becomes
\[
\frac{d^2v_{r}^{(1)}}{d\zeta^2}+\frac{\pi^2}{4}v_{r}^{(1)}=\frac{\pi\left(\xi+\frac{1}{\chi}\right)}{(1+\xi)}\left[\frac{1}{3}\cos\left(\frac{\pi \zeta}{2}\right)-\right.
\]
\begin{equation}\label{AA.9}
\left.-\frac{\pi}{4}\sin\left(\frac{\pi \zeta}{2}\right)\cos\left(\frac{\pi \zeta}{2}\right)\right]=0.
\end{equation}
This differential equation will have the solution
\[v_{r}^{(1)}(\zeta)=C_{1}\sin\left(\frac{\pi \zeta}{2}\right)+C_{2}\cos\left(\frac{\pi \zeta}{2}\right)\]
\begin{equation}\label{AA.10}
+\frac{\left(\xi+\frac{1}{\chi}\right)}{3 \pi}\frac{2\cos\left(\frac{\pi \zeta}{2}\right)+\pi \zeta   \sin\left(\frac{\pi \zeta}{2}\right)+\frac{\pi}{2}sin(\pi \zeta) }{(1+\xi)}.
\end{equation}
Applying the boundary condition $v_{r}^{(1)}(\zeta=1)=0$, we find the constant
\[C_{1}=-\frac{\left(\xi+\frac{1}{\chi}\right)}{3(1+\xi)}.\]
In order to find the value of $C_{2}$, we use the property of orthogonality, i.e.
\[
\int_{0}^{1}v_{r}^{(0)}v_{r}^{(1)}\;d\zeta=0,
\]
which result in
\[C_{2}=-\frac{7}{9}\frac{\left(\xi+\frac{1}{\chi}\right)}{(1+\xi)\pi}.
\]
As a result, the first order correction to $v_{r}$ corresponding to the fundamental mode is
\[v_{r}^{(1)}=\frac{\left(\xi+\frac{1}{\chi}\right)}{(1+\xi)}\left[-\frac{1}{3}\sin\left(\frac{\pi \zeta}{2}\right)-\frac{7}{9\pi}\cos\left(\frac{\pi \zeta}{2}\right)\right]+\]
\begin{equation}\label{AA.11}
+\frac{\left(\xi+\frac{1}{\chi}\right)}{(1+\xi)}\left[\frac{1}{3\pi}2\cos\left(\frac{\pi \zeta}{2}\right)+\pi \zeta    \sin\left(\frac{\pi \zeta}{2}\right)+\frac{\pi}{2}\sin(\pi \zeta) \right].
\end{equation}
In Figure A.1 we plot the correction to the eigenfunction for $\epsilon=0.1$, $\chi=0.9$, and $\xi=10$. Figure (\ref{fig9}) shows that we can approximate $v_{r}(z)$ by $\cos(\pi z/2L)$ since the first order correction brings changes of about  $1(\%)$, i.e. insignificant.

\subsection{Corrections to the first harmonic}
The same analysis can be repeated for the first harmonic, taking into account the right boundary conditions. After a straightforward calculation it is easy to show that
\begin{equation}\label{AA.13}
v_{r}^{(0)}(\zeta)=sin(\pi \zeta),
\end{equation}
\[v_{r}^{(1)}(\zeta)=\frac{\xi+1/\chi}{1+\xi}\cdot\left[-\frac{176}{225\pi}\sin(\pi \zeta)+\frac{4}{15}\cos(\pi \zeta)\right]\]
\[-\frac{16}{15\pi}\frac{\xi+1/\chi}{1+\xi}\left[\frac{5 \pi}{32}\cos\left (\frac{\pi \zeta}{2}\right)+\frac{3 \pi}{32}\cos\left(\frac{3 \pi \zeta}{2}\right)\right]\]
\begin{equation}\label{AA.14}
-\frac{16}{15\pi}\frac{\xi+1/\chi}{1+\xi}\left[\frac{\pi \zeta}{2}\cos(\pi \zeta)-\frac{1}{2}\sin(\pi \zeta)\right].
\end{equation}

The correction to the eigenfunction corresponding to the first harmonic has been plotted in Figure A.2 for the same values as before. It is obvious that the changes introduced by stratification in the value of the eigenfunction are of the order of
$2(\%)$, i.e. negligably small.
\begin{figure}
\centering
\includegraphics[width=\columnwidth]{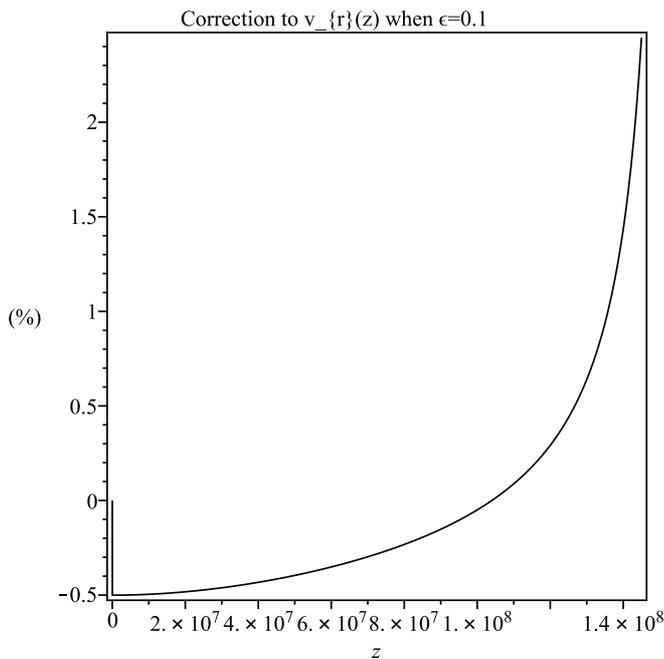}
\caption{The same as Figure A.1 but here we represent the correction to the eigenfunction for the first harmonic kink oscillation.}
\label{fig10}
\end{figure}
The two figures show that the effect of density stratification becomes more important for higher harmonics. Given the very large values of $\chi$ we used for prominences, the approximations used in this Appendix will always be valid. Fo the graphical represention of corrections in Figs A1 and A2 we used $\chi=1$. If we lower this value to, e.g. 0.7 the corrections would still be small since the maximum relative change in the eigenfunction describing the fundamental mode would be 1.1 \%, while for the first harmonic, this would increase to 3.5 \%.

The robustness of our analysis was checked using a full numerical investigation for arbitrary values of $\chi$ and $y$. A typical dependence of the $P_1/P_2$ period ratio with respect to $L/\pi H_i$ for one value of $\chi$ is shown in Fig. A.3 where the solid line corresponds to the analytical and the dotted line represent the numerical results, in both cases the density is inhomogeneous with respect to the coordinate $z$. The loop is set into motion using a gaussian-shaped source and we use a full reflective boundary conditions at the two footpoints of the loop. After the oscillations are formed, we use the FFT procedure to obtain the values of periods. Our analysis shows that the differences between the results obtained using the variational method and a full numerical investigation are of the order of 7\% but towards the large range of $L/\pi H_i$. Restricting ourself to realistic values, i.e. $L/\pi H_i<5$, we see that the results obtained with the two methods coincide with great accuracy.
\begin{figure}
\centering
\includegraphics[width=\columnwidth]{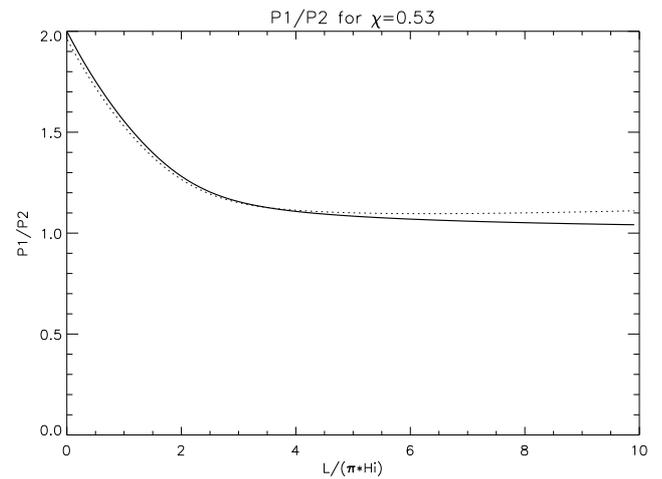}
\caption{Comparison of the analytical (solid line) and numerical (dotted line) results for the $P_1/P_2$ variation with $L/\pi H_i$ for coronal case corresponding to $\chi=0.53$. }
\label{fig11}
\end{figure}

\begin{thebibliography}{}
\bibitem[Andries et~al.(2005)]{andries2005}
Andries, J., Arregui, I., \& Goossens, M., 2005, \apjl, 624, L57
\bibitem[Andries et~al.(2009)]{andries2009}
Andries, J., van Doorsselaere, T., Roberts, B., Verth, G., Verwichte, E., \& Erd{\'e}lyi, R., 2009, \ssr, 149, 3
\bibitem[Aschwanden et~al.(2002)]{asch02}
Aschwanden, M.~J., de Pontieu, B., Schrijver, C.~J., \& Title, A.~M., 2002, \solphys, 206, 99
\bibitem[Aschwanden and Schrijver.(2011)]{AschwandenSchrijver}
Aschwanden, M.~J.., \& Schrijver, C. J., 2011, \apj, 736, 102
\bibitem[Ballai et~al.(2011)]{ballai2011}
Ballai, I., Jess, D., \& Douglas, M., 2011, \aap, 534, 13
\bibitem[Ballai et~al.(2005)]{ballai2005}
Ballai, I., Erd{\'e}lyi, R., \& Pint{\'e}r, B., 2005, \apjl, 633, L145
\bibitem[Ballai(2007)]{ballai2007}
Ballai, I., 2007, \solphys, 246, 177
\bibitem[Ballai et~al.(2008)]{ballai2008}
Ballai, I., Douglas, M., \& Marcu, A., 2008, \aap, 488, 1125
\bibitem[Ballai et~al.(2006)]{ballai2006}
Ballai, I., Erd\'elyi, R. \& Hargreaves, J., 2006, Phys. Plasmas, 14, 042108
\bibitem[Banerjee et~al.(2007)]{banerjee2007}
Banerjee, D., Erd{\'e}lyi, R., Oliver, R., \& O'Shea, E., 2007, \solphys, 246, 3
\bibitem[Berger et~al.(2011)]{berger2011}
Berger, T., Testa, P., Hillier, A. et al., 2011, { Nature}, 472, 197
\bibitem[De Moortel \& Brady(2007)]{DeMoortelBrady2007}
De Moortel, I., \& Brady, C.~S., 2007, \apj, 664, 1210
\bibitem[Dymova \& Ruderman (2005)]{dymova2005}
Dymova, M.~V. \& Ruderman, M.~S., 2005, \solphys 229, 79
\bibitem[Dymova \& Ruderman (2006)]{dymova2006}
Dymova, M.~V. \& Ruderman, M.~S., 2006, \aap 457, 1069
\bibitem[Edwin \& Roberts(1983)]{edwinroberts1983}
Edwin, P.~M., \& Roberts, B., 1983, \solphys, 88, 179
\bibitem[Erdelyi \& Ballai(2007)]{erdbal2007}
Erd\'elyi, R., \& Ballai, I., 2007, { Astron. Nachr.}, 328, 726
\bibitem[Eto et~al.(2002)]{eto2002}
Eto, S., et al., 2002, {\it PASJ}, 54, 481
\bibitem[Gruszecki et~al. (2006)]{gruszecki2006}
Gruszecki, M., Murawski, K., Selwa, M., Ofman, L., 2006, \aap,
460, 887
\bibitem[Gruszecki et~al. (2007)]{gruszecki2007}
Gruszecki, M., Murawski, K., Solanski, S.K., Ofman, L., 2007,
\aap, 469, 1117
\bibitem[Gruszecki et~al. (2008)]{gruszecki2008}
Gruszecki, M., Murawski, K., Ofman, L., 2008, \aap, 488, 757
\bibitem[Isobe \& Tripathi(2006)]{isobe08}
Isobe, H., \& Tripathi, D., 2007, \aap, 449, L17
\bibitem[Jing et~al.(2003)]{jing03}
Jing, J. et al., 2003, \apj, 584, L103
\bibitem[Klimchuk (2006)]{klim06}
Klimchuk., 2006, \solphys, 234, 41
\bibitem[Liu et~al.(2010)]{liu2010}
Liu, Wei, Nitta, Nariaki V., Schrijver, Carolus J., Title, Alan M., Tarbell, Theodore D., 2010, \apj, 723, 53
\bibitem[McEwan et~al.(2006)]{mcewan2006}
McEwan, M.~P., D{\'{\i}}az, A.~J., \& Roberts, B., 2008, \aap, 481, 819
\bibitem[McEwan et~al.(2006)]{mcewan2006.1}
McEwan, M.~P., Donnelly, G.~R., D{\'{\i}}az, A.~J., \& Roberts, B., 2006, \aap, 460, 893
\bibitem[McLaughlin \& Ofman(2008)]{mclaughlin2008}
McLaughlin, J.A., Ofman, L., 2008, \apj, 682, 1338
\bibitem[Moreton \& Ramsey(1960)]{moreton1960}
Moreton, G.E., Ramsey, H.E., 1960, { Astron. Soc.}, 72,
357
\bibitem[Morton, Ruderman \& Erdelyi(2009)]{morton11a}
Morton, R., Ruderman, M.S. \& Erd\'elyi, R., 2011, \aap, 534, 27
\bibitem[Morton \& Ruderman(2011)]{morton11b}
Morton, R. \& Ruderman, M.S., 2011, \aap, 527, 53
\bibitem[Morton \& Erdelyi(2009)]{morton09}
Morton, R. \& Erd\'elyi, R., 2009, \aap, 502, 315
\bibitem[Mulu-Moore et~al.(2011)]{mulu2011}
Mulu-Moore, F.M., Winebarger, A. R., Warren, H.~P., \& Aschwanden, M.J, 2011, \apj, 733, 59
\bibitem[Murawski \& Misielak(2010)]{mur10}
Murawski, K. \& Musielak, Z. E., 2010, \aap, 518, 37
\bibitem[Nakariakov et~al.(1999)]{nakariakov1999}
Nakariakov, V.~M., Ofman, L., Deluca, E.~E., Roberts, B., \& Davila, J.~M., 1999, Science, 285, 862
\bibitem[Nakariakov \& Ofman(2001)]{nakariakov2001}
Nakariakov, V.~M., Ofman, L., 2001, \aap, 372, 253
\bibitem[Ofman \& Aschwanden (2002)]{ofm02a}
Ofman, L., Aschwanden, M.J., 2002, \apj, 576, L153-L156
\bibitem[Ofman \& Thompson(2002)]{ofm02}
Ofman, L., Thompson, B.J., 2002, \apj, 574, 440
\bibitem[Ofman(2007)]{ofm07}
Ofman, L., 2007, \apj, 665, 1134
\bibitem[Ofman(2009)]{ofm09}
Ofman, L., 2009, \apj, 694, 502
\bibitem[Ofman \& Wang(2008)]{ofm08}
Ofman, L. \& Wang, T. J., 2008, \aap, 482, 9
\bibitem[Okamoto et~al.(2004)]{okamoto2004}
Okamoto, T. J., Nakai, H., \& Keiyama, A., 2004, \apj, 608, 1124
\bibitem[Patsourakos \& Vourlidas(2009)]{patsourakos09}
Patsourakos, Spiros, Vourlidas, Angelos, 2009, \apj Letters, 700,
L182
\bibitem[Pinter et~al.(2007)]{pinter2008}
Pint\'er, B., Jain, R., Tripathi, D., \& Isobe, H., 2008, \apj, 680, 1560
\bibitem[Rae \& Roberts(1982)]{rae82}
Rae, I. C. \& Roberts, B., 1982, \apj, 256, 761
\bibitem[Ramsey and Smith (1966)]{ramsey66}
Ramsey, H.E., \& Smith, S.~F., 1966, {\it AJ}, 71, 197
\bibitem[Roberts et~al.(1984)]{roberts1984}
Roberts, B., Edwin, P.~M., \& Benz, A.~O., 1984, \apj, 279, 857
\bibitem[Ruderman \& Erd{\'e}lyi(2009)]{RudermanErdelyi2009}
Ruderman, M.~S., \& Erd{\'e}lyi, R., 2009, \ssr, 149, 199
\bibitem[Ruderman \& Roberts(2002)]{ruderman02}
Ruderman, M.~S., \& Roberts, B., 2002, \apj, 577, 475
\bibitem[Ruderman et~al.(2008)]{ruderman2008}
Ruderman, M.~S., Verth, G., \& Erd{\'e}lyi, R., 2008, \apj, 686, 694
\bibitem[Ruderman(2011)]{ruderman2011}
Ruderman, M.~S., 2011, \solphys, 271, 41
\bibitem[Selwa \& Ofman(2009)]{selwa09}
Selwa, M., Ofman, L., 2009, {\it Ann. Geophys.}, 27, 3899
\bibitem[Selwa \& Ofman(2010)]{selwa10}
Selwa, M., Ofman, L., 2010, \apj, 714, 170
\bibitem[Selwa et~al.(2010)]{selwa10a}
Selwa, M., Murawski, K., Solanki, S.~K., \& Ofman, L., 2010, \aap, 512, A76
\bibitem[Selwa et~al.(2011)]{selwa2011}
Selwa, M., Ofman, L., Solanski, S.K., 2011, \apj, 726, 42
\bibitem[Selwa et~al.(2011)]{selwa11}
Selwa, M., Solanski, S.K., Ofman, L., 2011, \apj, 728, 87
\bibitem[Thompson et-al.(1999)]{thompson1999}
Thompson, B.J., Gurman, J.B., Neupert, W.M., Newmark, J.S.,
Dellaboudini\'{e}re, J.-P., St., Gyr, O.C., Stezelberger, S.,
Dere, K.P., Howard, R.A., Michels, D.J., 1999, \apj, 517, L151
\bibitem[Uchida(1970)]{uchida1970}
Uchida, Y., 1970, {Astron Soc. Japan}, 22, 341
\bibitem[Van Doorsselaere et~al.(2009)]{vandoorsselaere2009}
Van Doorsselaere, T., Birtlill, D,C,C., \& Evans, G. R., 2009, \aap, 508, 1485
\bibitem[Van Doorsselaere et~al.(2007)]{vandoorsselaere2007}
Van Doorsselaere, T., Nakariakov, V.~M., \& Verwichte, E., 2007, \aap, 473, 959
\bibitem[Van Doorsselaere et~al.(2008)]{vandoors08}
Van Doorsselaere, T., Ruderman, M.S., \& Robertson, D., 2008, \aap, 485, 849
\bibitem[Verth et~al.(2007)]{verth2007}
Verth, G., Van Doorsselaere, T., Erd{\'e}lyi, R., \& Goossens, M., 2007, \aap, 475, 341
\bibitem[Verth \& Erd{\'e}lyi(2008)]{VerthErdelyi2008}
Verth, G., \& Erd{\'e}lyi, R.\ 2008, \aap, 486, 1015
\bibitem[Verth et~al.(2008)]{VerthErdelyiJess2008}
Verth, G., Erd{\'e}lyi, R., \& Jess, D.~B., 2008, \apjl, 687, L45
\bibitem[Verwichte et~al.(2009)]{verwichte2009}
Verwichte, E., Aschwanden, M.J., Van Doorsselaere, T., Foullon, C, \& Nakariakov, V.~M., 2009, \apj, 698, 397
\bibitem[Warren et~al.(2008)]{warren2008}
Warren, H. P., Ugarte-Urra, I., Doschek, G.A. et al., 2008, \apj, 686, 131
\bibitem[Wang et~al. (2003a)]{wang2003}
Wang, T. J., Solanki, S. K., Curdt, W., Innes, D. E., Dammasch, I. E., \& Kliem, B. 2003a, \aap, 406, 1105
\bibitem[Wills-Davey and Thompson(1999)]{wills1999}
Wills-Davey, M.J., Thompson, B.J., 1999, \solphys, 190, 467
\bibitem[Winebarger et~al.(2003)]{winebarger2003}
Winebarger, A. R., Warren, H.~P., \& Seaton, D.B, 2003, \apj, 593, 1164
\end{thebibliography}
\end{document}